\documentclass[10pt]{iopart}

\usepackage{iopams}
\usepackage{cite}
\usepackage{array}
\usepackage{amssymb}
\usepackage[cp1251]{inputenc}
\usepackage[T2A]{fontenc}
\usepackage{graphicx}
\usepackage[dvipsnames]{color}
\usepackage{colortbl}
\usepackage{multirow}
\usepackage{pdfsync}
\usepackage{footnote}
\usepackage{threeparttable}
\usepackage{braket}
\usepackage[english]{babel}
\begin{document}
\newcolumntype{-}{@{}}
\newcolumntype{C}{>{\displaystyle}c}
\newcolumntype{L}{>{\displaystyle}l}
\newcolumntype{R}{>{\displaystyle}r}
\newcolumntype{d}[1]{D{,}{,}{#1}}
\newcolumntype{=}{@{\,}C@{\,}}
\newcommand{\red}{\textcolor{black}}
\newcommand{\MC}{\color{blue}}
\renewcommand{\thefootnote}{\arabic{footnote}}

\title{Separated Schmidt modes in the angular spectrum of biphotons}

\author{
N.~A.~Borshchevskaia$^{1}$, F.~Just$^{2}$, K.~G.~Katamadze$^{1,3,4}$,
A.~Cavanna$^{2,5}$, and M.~V.~Chekhova$^{2,5}$}

\address{
\emph{
$^1$ Quantum Technology Centre, Faculty of Physics, M. V. Lomonosov Moscow State University, Leninskie Gory 1, 119991 Moscow, Russia\\
$^2$ Max Planck Institute for the Science of Light, Staudtstrasse 2, 91058 Erlangen, Germany\\
$^3$ Institute of Physics and Technology, Russian Academy of Sciences, Nakhimovsky prospect 36, 117218 Moscow, Russia\\
$^4$ National Research Nuclear University MEPhI, Kashirskoe Shosse 31, 115409 Moscow, Russia\\
$^5$ University of Erlangen-Nuremberg, Staudtstrasse 2, 91058 Erlangen, Germany}}

\ead{borschxyz@gmail.com}

\begin{abstract}
We prepare qudits based on angular multimode biphoton states by modulating the pump angular spectrum. The modes are prepared in the Schmidt basis and their intensity distributions do not overlap in space. This allows one to get rid of filtering losses while addressing single modes and to realize a single-shot qudit readout.
\end{abstract}

\vspace{2pc}
\noindent{\it Keywords}: spontaneous parametric down-conversion, angular spectrum, qudit,  Schmidt decomposition
%
%
%
\\

\ioptwocol

\section{Introduction}

Spontaneous parametric downconversion (SPDC) is a simple high-rate and high-fidelity source of entangled quantum states. Entangled polarization qubits is an ideal system for demonstration of quantum teleportation, quantum dense coding, quantum key distribution etc. But for further improvement of listed techniques, it is necessary to increase the system dimensionality and turn to entangled qudits with much higher degree of entanglement  \cite{stenholm1998,Wang2005,Zhao2012a,Bourennane2001,Bruss2002,Groblacher2006}.

\begin{figure}[]
  \begin{center}
  \includegraphics[width=\linewidth]{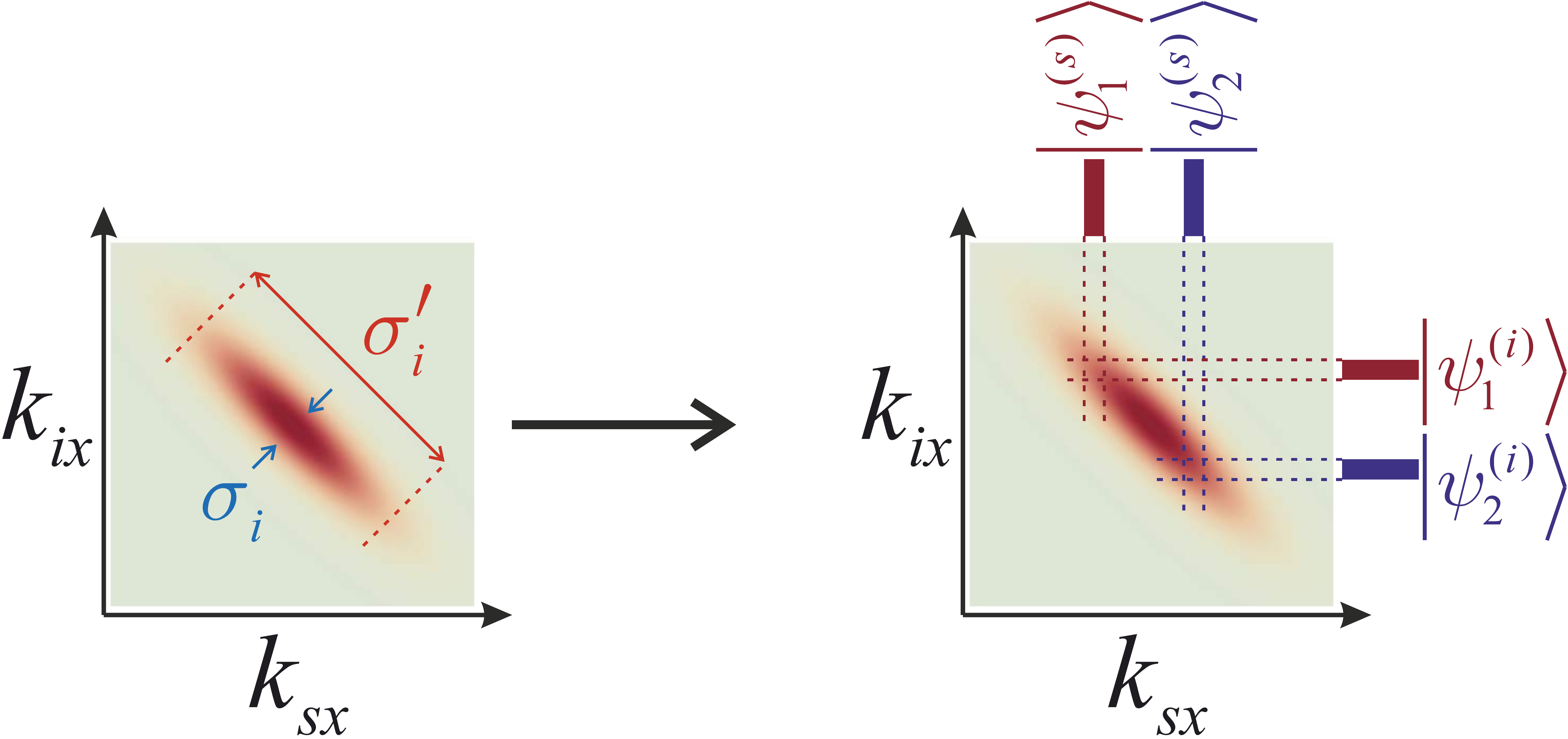}
  \caption{(a) Schematic illustration of the double-Gaussian TPA~\cite{Fedorov2009} and (b) the modes $f_j^{(s),(i)}$  obtained after slits placed in the far field into the signal and idler beams.}
  \label{fig:two-gaussian model}
  \end{center}
  \end{figure}

For this purpose one can exploit frequency \cite{Avella2014}, temporal \cite{Franson1989,Bechmann-Pasquinucci2000a,th2004,DeRiedmatten2004,Stucki2005}, and spatial SPDC modes \cite{OSullivan-Hale2005,Neves2005,Bartuskova2006,Walborn2006,Rossi2009}, including orbital angular momentum \cite{Vaziri2002,Malik2014} modes. Frequency and temporal modes do not allow one to manipulate and register them without postselection \cite{Gisin2002}. This can be overcome by using spatial modes and  integrated optics \cite{Harris2017}.
The biphoton quantum state generated  through spontaneous parametric down-conversion (SPDC) can be written as
\begin{equation}\label{Monken1}
\begin{array}{l}
\ket{\Psi} =  \mbox{const}\times
\int d\mathbf{k}_s \int d\mathbf{k}_i \, 
F(\mathbf{k}_s,\mathbf{k}_i)  \ket{1_{{k}_s},1_{{k}_i}} \\
\phantom{asfhnhsgfbsbsbgbtnsrncvnvvdgdbbgbfd} + \ket{vac},
\end{array}
\end{equation}
where signal, idler and pump photons are characterised by wavevectors $\mathbf{k}_s$, $\mathbf{k}_i$, $\mathbf{k}_p$, $\ket{1_{{k}_s},1_{{k}_i}}$ denotes a pair of signal and idler photons with wavevectors $\mathbf{k}_s$, $\mathbf{k}_i$,
and $F(\mathbf{k}_s,\mathbf{k}_i)$ is the two-photon amplitude (TPA)~\cite{Monken1998}. If the pump has a narrow transverse wavevector spectrum $v(\mathbf{k}_{p\perp})$ so that $\mathbf{k}_{p\perp}\ll\mathbf{k}_{p}$, the TPA has the form
\begin{equation}\label{Monken2}
\begin{array}{l}
F(\mathbf{k}_s,\mathbf{k}_i)=v(\mathbf{k}_{s\perp}+\mathbf{k}_{i\perp}) 
\mbox{sinc}\left[\frac{L}{2}\Delta k_z
\right],
\end{array}
\end{equation}
where $\perp$ denotes the transverse component of the wavevector, $\Delta k_z$ is the component of the wavevector mismatch $\mathbf{k}_p-\mathbf{k}_s-\mathbf{k}_i$ parallel to the pump wavevector, and $L$ is the length of the crystal. Here the sinc function can be approximated by a Gaussian function as proposed in Ref.~\cite{Banaszek2003}:
\begin{equation}\label{Fedorov}
\begin{array}{l}
F(\mathbf{k}_s,\mathbf{k}_i)=
 \exp\left(-\frac{( \mathbf{k}_s+\mathbf{k}_i)^2}{2\sigma_k^2}\right) \exp\left(-\frac{( \mathbf{k}_s-\mathbf{k}_i)^2}{2\sigma_k^{\prime 2}}\right) ,
\end{array}
\end{equation}
where $\sigma_k$ reflects the pump angular divergence and $\sigma_k^{\prime}$ is determined by phase matching relations (Fig.~\ref{fig:two-gaussian model}a). For type-I collinear and non-collinear frequency-degenerate SPDC, $\sigma_k^{\prime}$ takes the values $\sqrt{4k_p/(\gamma_1 L)}$~\cite{law2004} and $\sqrt{n_s}/(L\sqrt{(n_s-n_p) \gamma_2})$~\cite{Fedorov2018,Fedorov2015a}, respectively, where $n_s$ and $n_p$ denote the signal and pump refractive indices, and $\gamma_1=0.249$ \cite{Fedorov2009} and $\gamma_2=0.195$ \cite{Fedorov2015a} are the parameters which come from the approximations $\mbox{sinc}(x^2) \approx \exp{(-\gamma_1 x^2)}$ and $\mbox{sinc}(x) \approx \exp{(-\gamma_2 x^2)}$.

  \begin{figure}[]
  \begin{center}
  \includegraphics[width=\linewidth]{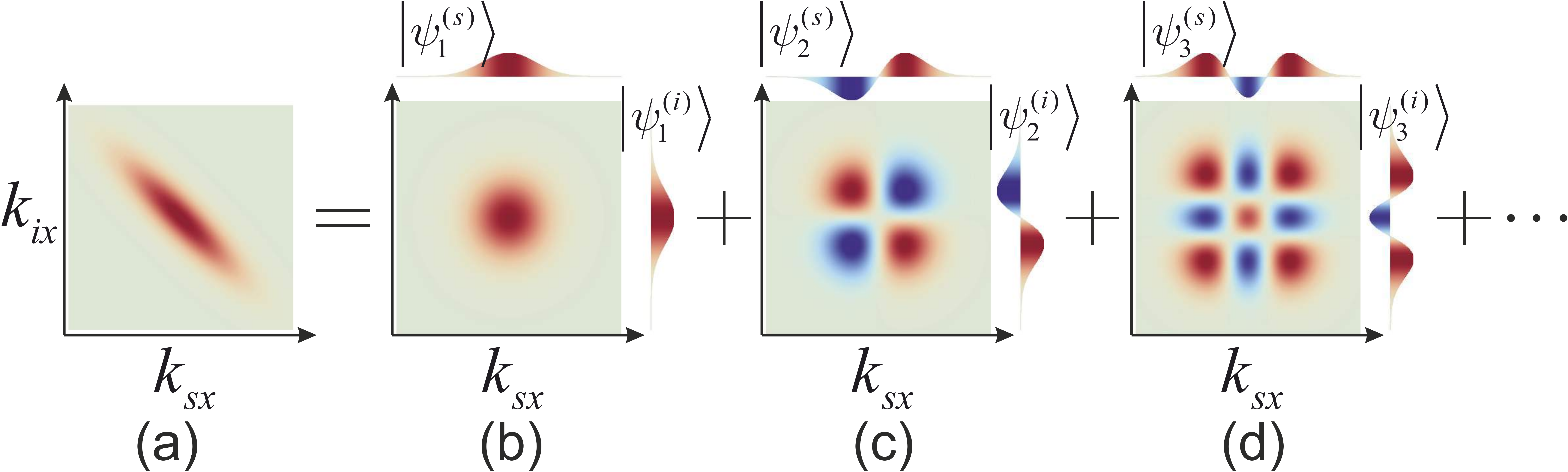}
  \caption{First terms (b-d) in Hermite-Gauss decomposition of the double-Gaussian TPA (a).}
  \label{fig:HG}
  \end{center}
  \end{figure}

The TPA of two entangled qudits, each having $d$ modes, has the form of a single sum
\begin{equation}\label{qudit}
\begin{array}{l}
\ket{\Psi} =  \sum\limits_{m=1}^{d} c_m \ket{\psi_m^{(i)}} \ket{\psi_m^{(s)}},
\end{array}
\end{equation}
where $\left\{ \psi_m \right\}$ denotes a set of eigenvectors of the reduced density matrices for the signal and idler photons and $|c_m|^2$ is the probability of registering the state in the $m$th mode.
To extract entangled modes in signal and idler channels one often places slits in front of the detector (Fig.~\ref{fig:two-gaussian model}b). The thinner the slits, the closer the extracted state to a single-mode one, but, at the same time, the higher the losses and the portion of uncorrelated photons in each mode.

A more useful approach realized here is to choose qudit modes to coincide with Schmidt modes \cite{law2004}. To achieve this we should perform decomposition of TPA in the form of a single sum of factorized terms instead of double integration (\ref{Monken1}):
\begin{equation}\label{TPA_qudit}
\begin{array}{l}
F(\mathbf{k}_s,\mathbf{k}_i)=  \sum\limits_{m=1}^{d} c_m f^{(s)}_m(\mathbf{k}_s) f^{(i)}_m(\mathbf{k}_i),
\end{array}
\end{equation}
with the profiles of the Schmidt modes $f^{(s,i)}_m(\mathbf{k}_{s,i})$ non-overlapping in the $\mathbf{k}$ space.

\begin{figure}[]
  \begin{center}
  \includegraphics[width=\linewidth]{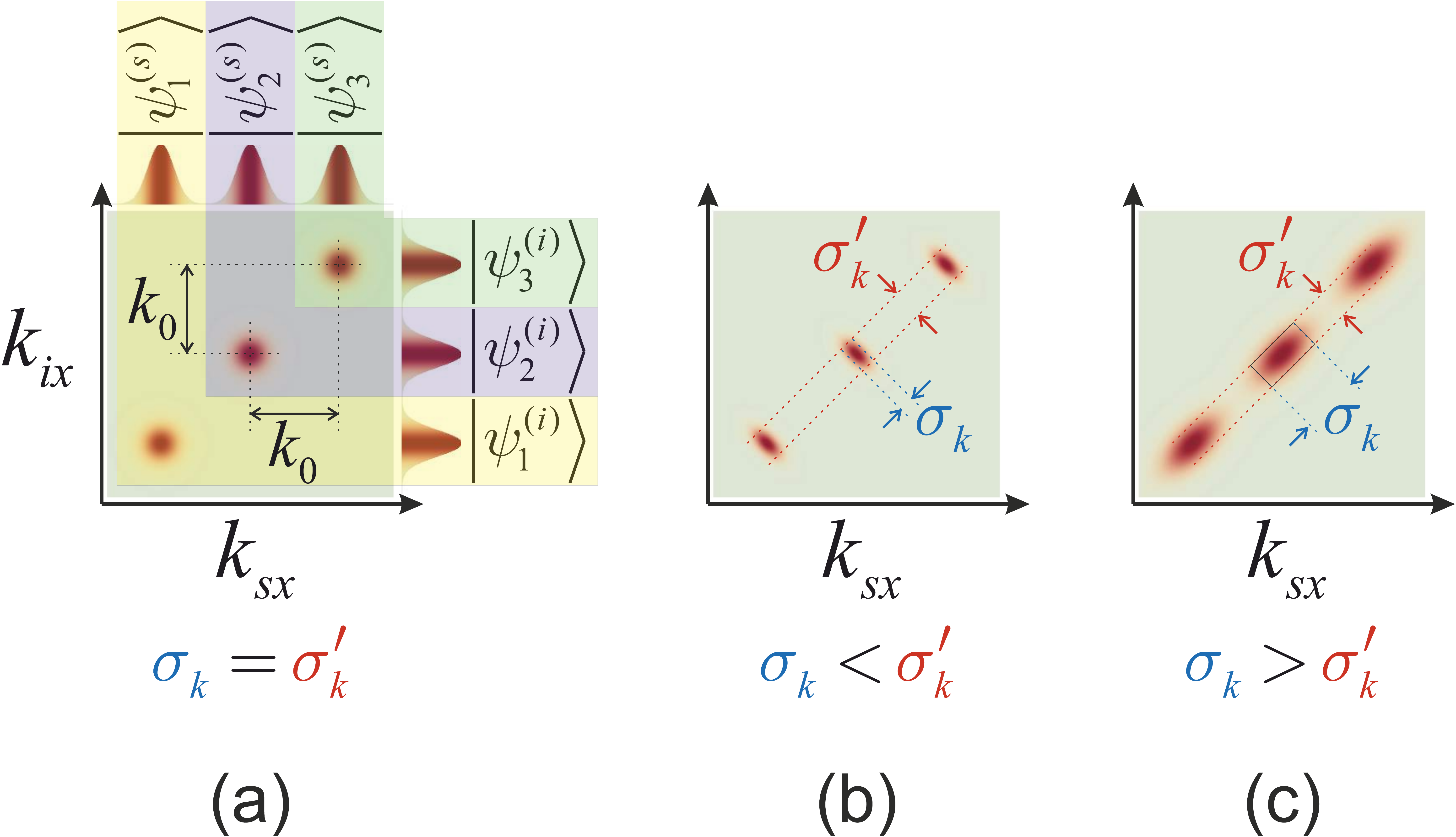}
  \caption{The TPA angular distribution for different pump beam divergency.}
  \label{fig:pump_round_elliptical}
  \end{center}
  \end{figure}

\begin{figure}[]
  \begin{center}
  \includegraphics[scale=0.45]{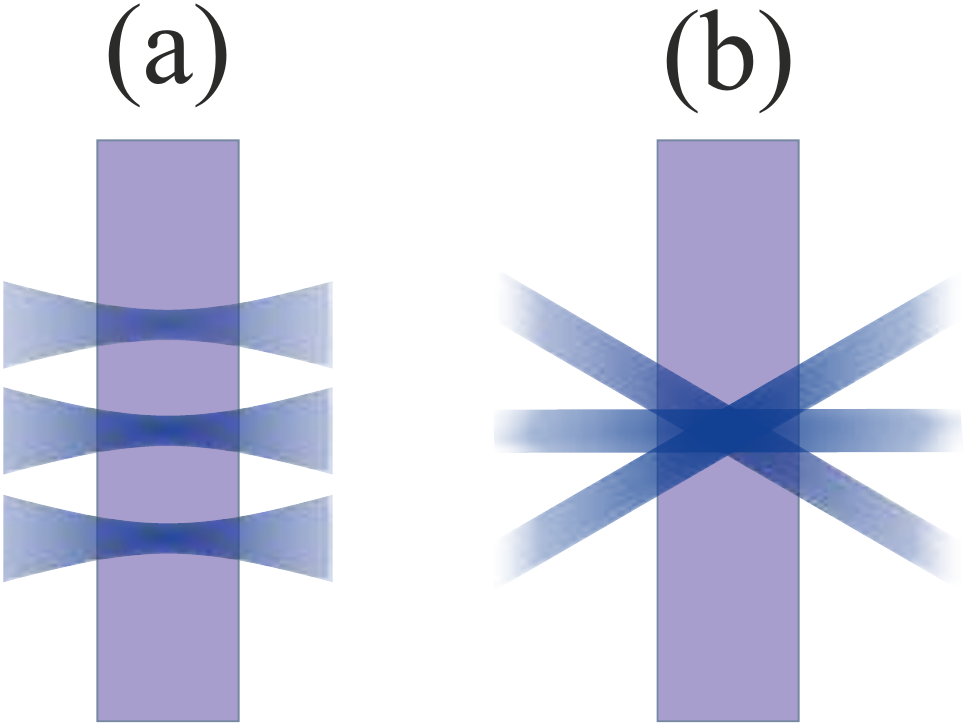}
  \caption{Realisation of separated modes in the near (a) or far field (b).}
  \label{fig:focusing_types}
  \end{center}
  \end{figure}

For a double-Gaussian TPA, the Schmidt modes can be chosen in the Hermite-Gauss or Laguerre-Gauss basis \cite{Straupe2011} and will spatially overlap (Fig.~\ref{fig:HG}). Projective measurements in this case require a spatial light modulator (SLM) for each mode combined with a spatial filter (e.g. single-mode fiber), which is an origin of considerable losses and does not allow a single-shot qudit readout. To overcome it one has to use complicated mode-sorting schemes based on the phase modulation \cite{Zhou2017,Ruffato2018, Zhou2019} which also leads to the drop of the efficiency. Non-overlapping Schmidt modes can help to eliminate this disadvantage (Fig.~\ref{fig:pump_round_elliptical}b). Similarly to how it was demonstrated in a frequency domain \cite{Avella2014}, it can be realized for spatial mode in the near or far field (Fig.~\ref{fig:focusing_types}).

Here we demonstrate the preparation of Gaussian-shaped spatially separated Schmidt modes in the far field. Our approach is based on the TPA dependence on the transverse pump beam shape (\ref{Monken2}). 
In the three-mode SPDC state demonstrated in our work, each mode is generated from the corresponding Gaussian pump beam, coming to the crystal at a slightly different angle (Fig.~\ref{fig:pump_round_elliptical},~\ref{fig:focusing_types}b).
Note that the divergence of each pump beam should be adjusted in such a way that each maximum of the TPA shows no wavevector correlations and can be therefore treated as a single term in the Schmidt decomposition (5) (Fig.~\ref{fig:pump_round_elliptical}). The required divergence depends on the phasematching conditions for a given crystal. In the recent work \cite{Ghosh2018} the authors used a similar approach of pump beam modulation to achieve the right pump profile on the crystal surface. In contrast to current work they placed slits in the near field of the pump beam (Fig.~\ref{fig:focusing_types}a) and didn't exploit the Schmidt decomposition framework. Because of that the width of the slits was about 3 times higher then the value determined from the expression below (\ref{Fedorov}).

\section{Theoretical description}
To prove the principle we chose the type-I degenerate noncollinear SPDC generation regime characterized by the transverse wavevector $K$
and the far-field biphoton distribution to be a set of $M$ separated Gaussian peaks along the $x$ axis (perpendicular to the pump wavevector, see Fig.~\ref{fig:setup}):
\begin{equation} \label{PDCshape}
	\begin{array}{l}
F(k_{s_x},k_{i_x})=
\sum\limits_{m=0}^{M-1} \exp\left(-\frac{( k_{s_x}+k_{i_x}-k(m))^2}{2\sigma_k^2}\right) \\
\times \left[ \exp\left(-\frac{( k_{s_x}-k_{i_x}-K)^2}{2\sigma_k^{\prime 2}}\right)
+\exp\left(-\frac{( k_{s_x}-k_{i_x}+K)^2}{2\sigma_k^{\prime 2}}\right) \right] ,
	\end{array}
\end{equation}
where $K=2\pi \sqrt{2n_s(n_s-n_p)}/\lambda_p=k_s \sin(\theta_{s0})$ \cite{Fedorov2015a}, $\lambda_p$ is the pump wavelength, $k(m)=(M-1-2m)k_0/2$, $k_0$ defines the distance between neighbouring peaks (Fig.~\ref{fig:pump_round_elliptical}a), and $\theta_{s0}$ is the angle between the pump and signal beams at the degenerate wavelength.  If $\sigma_k=\sigma_k^{\prime}$, (\ref{PDCshape}) becomes factorized and acquires the form of the Schmidt decomposition.

The distribution of the pump field along the $x$ coordinate on the crystall surface is a product of a periodic function, which defines the distance between the neighboring biphoton Schmidt modes in the far field, and a Gaussian envelope, which defines the angular size of each mode:
 \begin{equation}\label{E_p}
	E_p(x) \propto
	e^{-\frac{ x^2 \sigma_k^2}{2}}\sum\limits_{m=0}^{M-1} \cos(x k(m) )
\end{equation}

 Consequently, to obtain a set (\ref{PDCshape}) of separated Schmidt modes in the angular spectrum of SPDC we should choose $\sigma_k$ so that without the pump cosine modulation a single-mode regime in SPDC spectrum is achieved. 


\section{Experiment}
\begin{figure}[]
  \begin{center}
  \includegraphics[scale=0.7]{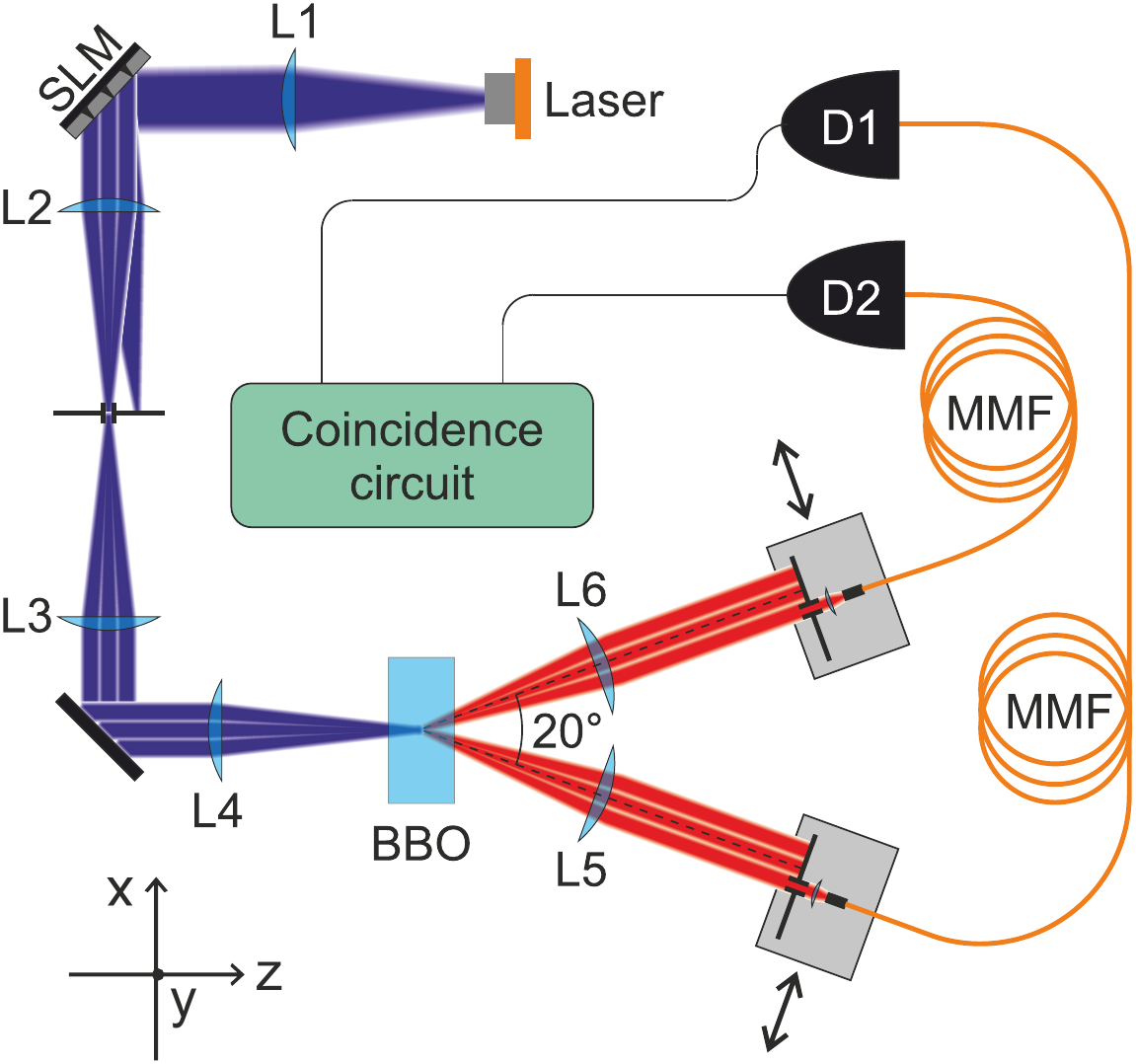}
  \caption{The experimental setup. The cw diode laser pump beam at 405~nm is expanded and reflected from the SLM. The first-order diffracted part is focused on the 3~mm BBO~crystal cut for type-I non-collinear phase matching therefore the pump phase modulation is performed in the far field.
  SPDC signal is registered also in the far field by scanning in the focal planes of lenses L5 and L6 spatial filters consisting of slits followed by multimode fibers.
}\label{fig:setup}
  \end{center}
  \end{figure}
Our setup is shown in Fig.~\ref{fig:setup}.
 The two-photon light was generated in noncollinear degenerate regime (the angle between the pump and signal beams at the degenerate wavelength outside the crystal was equal to $10^{\circ}$) in a 3~mm thick BBO crystal from a cw diode-laser pump with the wavelength $405\pm0.01$~nm.
 To achieve the desired pump distribution we used a spatial light modulator (SLM) Holoeye Pluto-VIS and prepared holograms according to the paper \cite{Bolduc2013}. The phase encrypted into the hologram (Fig.~\ref{fig:SLM}) had the form of a blazed diffraction grating (6 pixels of SLM per one period) with a three-peak Gaussian envelope defining pump amplitude modulation in the far field (\ref{PDCshape}).
  The pump beam was expanded before the SLM 
and its first-order diffracted part was focused on the crystal by lenses $L2,\, L3,\, L4$, and the zero-order diffracted part was filtered out. The pump power on the crystal was 0.5~mW.

The SPDC signal was registered in the far field by the
 spatial filters. Each filter consisted of a slit  placed in the focal plane of a collimating lens $L5 (L6)$, with the focal distance 100~mm, and followed by a lense of 2.1~mm focal distance and a multimode fiber. The filters were scanned in the focal planes of the lenses $L5 (L6)$.

The slits for spatial filtering had sizes of 0.2$\times$4~mm in one channel and 0.4$\times$10~mm in another one (the smaller size along the direction of the pump modulation).
For frequency filtration we placed in front of the detectors bandpass filters with the full width at half maximum (FWHM) 10~nm and central wavelength 810~nm. The fibers transmitted the coupled light to Perkin\&Elmer single-photon detectors based on avalanche photodiodes.

\section{Results and discussion}

\begin{figure}[t]
  \begin{center}
  \includegraphics[scale=0.3]{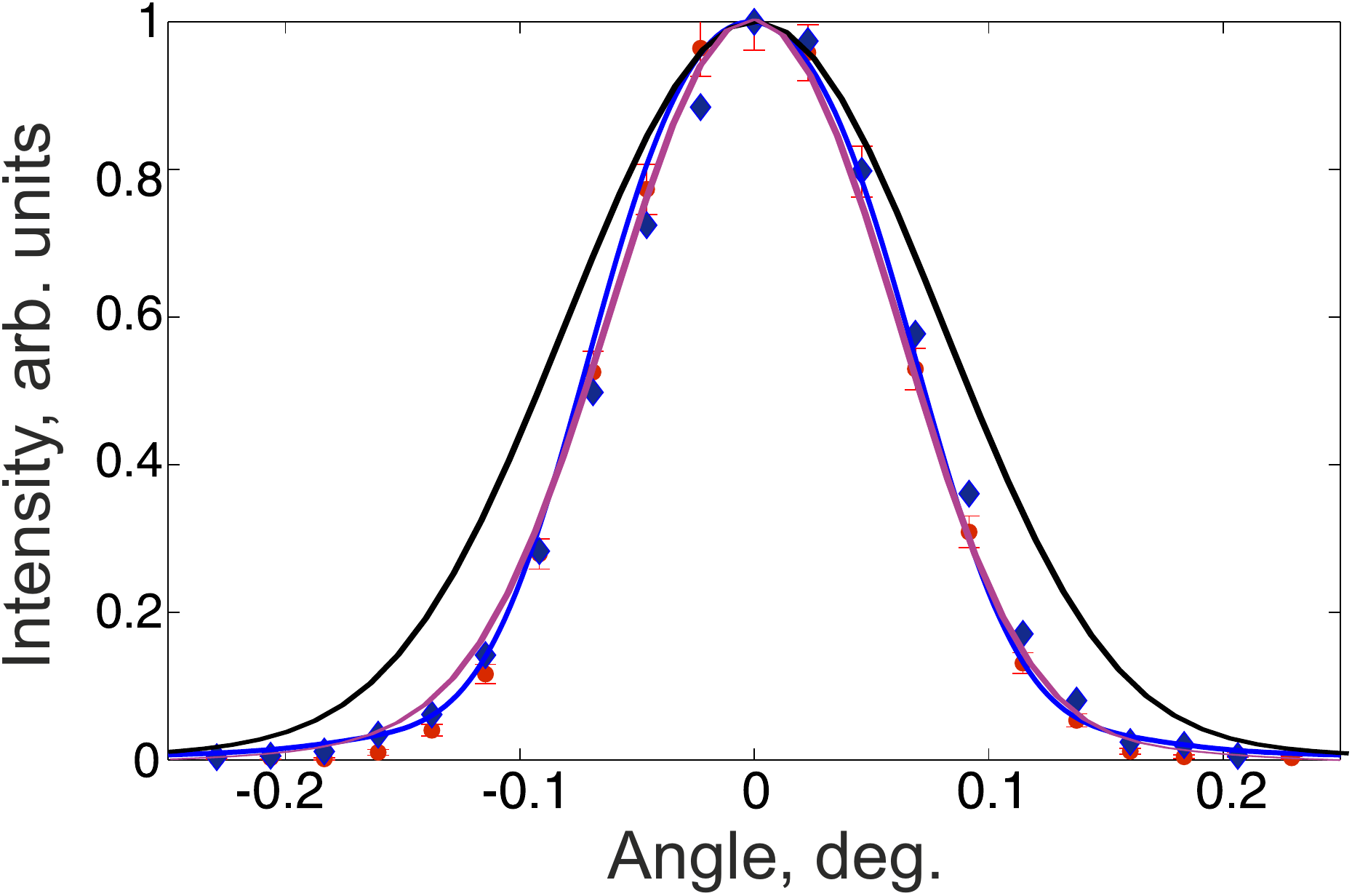} 
  \caption{Measured angular distribution of the SPDC signal for a non-modulated pump beam. Blue dots correspond to a single detector count rate, red dots to the coincidence count rate, the blue line is a Gaussian fit of the experimental data. Purple and black lines show theoretical predictions for the same pump beam size and SPDC only at degenerate wavelength or with the same spectral width as in experiment (10 nm), correspondingly.}\label{fig:1mode}
  \end{center}
  \end{figure}

At the first stage we adjusted the pump beam size in order to fulfill the equation $\sigma_k=\sigma_k^\prime$ (Fig.~\ref{fig:pump_round_elliptical}b) so that a single Gaussian pump beam would produce spatially single-mode biphotons. This condition was satisfied by using the pump with the FWHM   $(250\pm2)$~um.
The results are shown in Fig.~\ref{fig:1mode}.
 Blue dots correspond to the single detector counting rate, red ones show coincidences between the two detectors one of which was collecting all radiation in the signal channel and the second one was scanning along the modulation axis in the idler channel.
The ratio of the widths of these two curves is very close to the Schmidt number of the state \cite{Fedorov2004}. Thus, we can conclude that the state is nearly single-mode.

\begin{figure}[]
  \begin{center}
  \includegraphics[width=\linewidth]{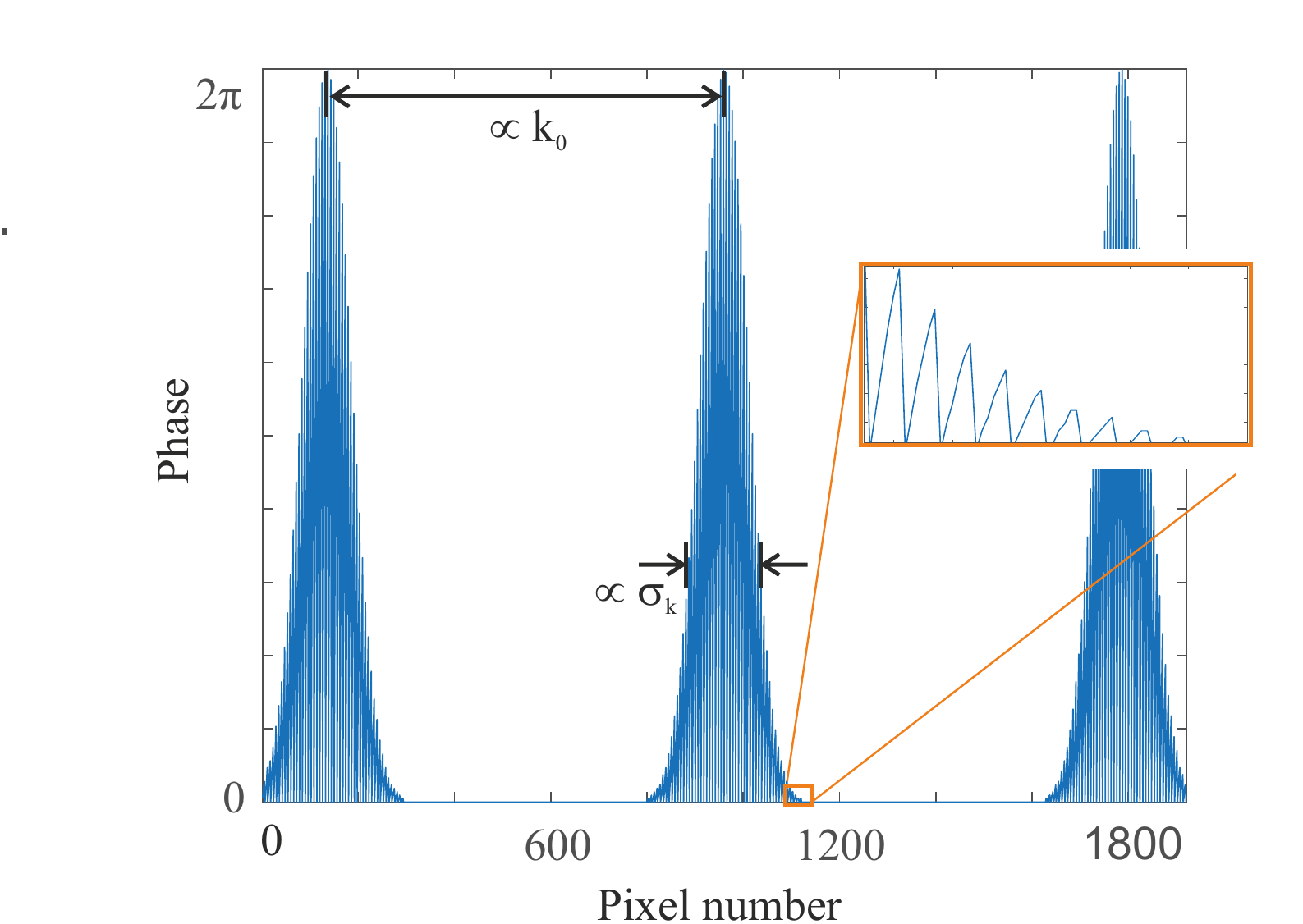} 
  \caption{The phase encrypted to the spatial light modulator in the $x$ direction.}\label{fig:SLM}
  \end{center}
  \end{figure}

\begin{figure}[b]
  \includegraphics[scale=0.42]{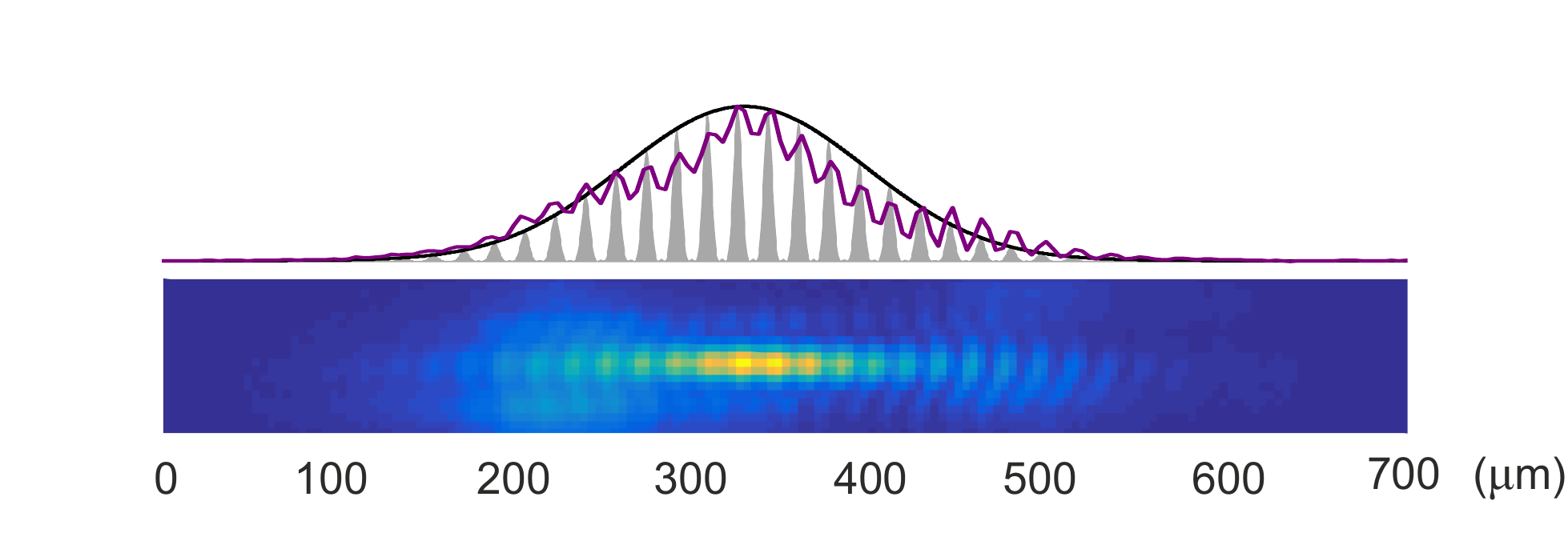}
  \caption{The pump intensity distribution obtained with a CCD camera placed instead of the BBO crystal and its central cross section (top, magenta line). The grey area shows the fit with Eq.~(\ref{alpha}), and the black line is its Gaussian envelope.}
  \label{fig:pump_view}
  \end{figure}

Next, we have modulated the angular distribution of the pump field according to (\ref{E_p}). The corresponding hologram sent to the SLM is shown in Fig.~\ref{fig:SLM}. As a result, we got spatially multimode SPDC generation with $M$=3. The control parameters were $k_0$ (inversely proportional to the period of the pump modulation at the crystal) and $\sigma_k$ (inversely proportional to the width of the envelope~$\delta x$). The distribution of the pump intensity on a CCD camera placed at the position of the crystal is shown in Fig.~\ref{fig:pump_view}.

Because the pump beam incident on the SLM had a Gaussian intensity distribution, the central peak in the resulting PDC spectrum was brighter than the sidebands. To reflect this fact we inserted a parameter $\alpha=0.63$ into Eq.~(\ref{E_p}) and took it into account in all theoretical calculations:
 \begin{equation}\label{alpha}
	E_p(x) \propto
	(1/2+  \alpha \cos(2k_0x)) \exp\left(-\frac{x^2 \sigma_k^2}{2}\right).
\end{equation}
	
Using the cross-section of the CCD image we estimated the parameters of the pump distribution: the field envelope FWHM was equal to $(246\pm2)\,\mu{\rm m}$ and $k_0=(0.168\pm0.002)\,\mu{\rm m}^{-1}$.
Note that the resolution of the camera (the size of each pixel 4.4~$\mu{\rm m}$) was not good enough to properly resolve the peaks but still allowed us to estimate the period and the width of the envelope.

\begin{figure}[]
  \begin{center}
  \includegraphics[scale=0.35]{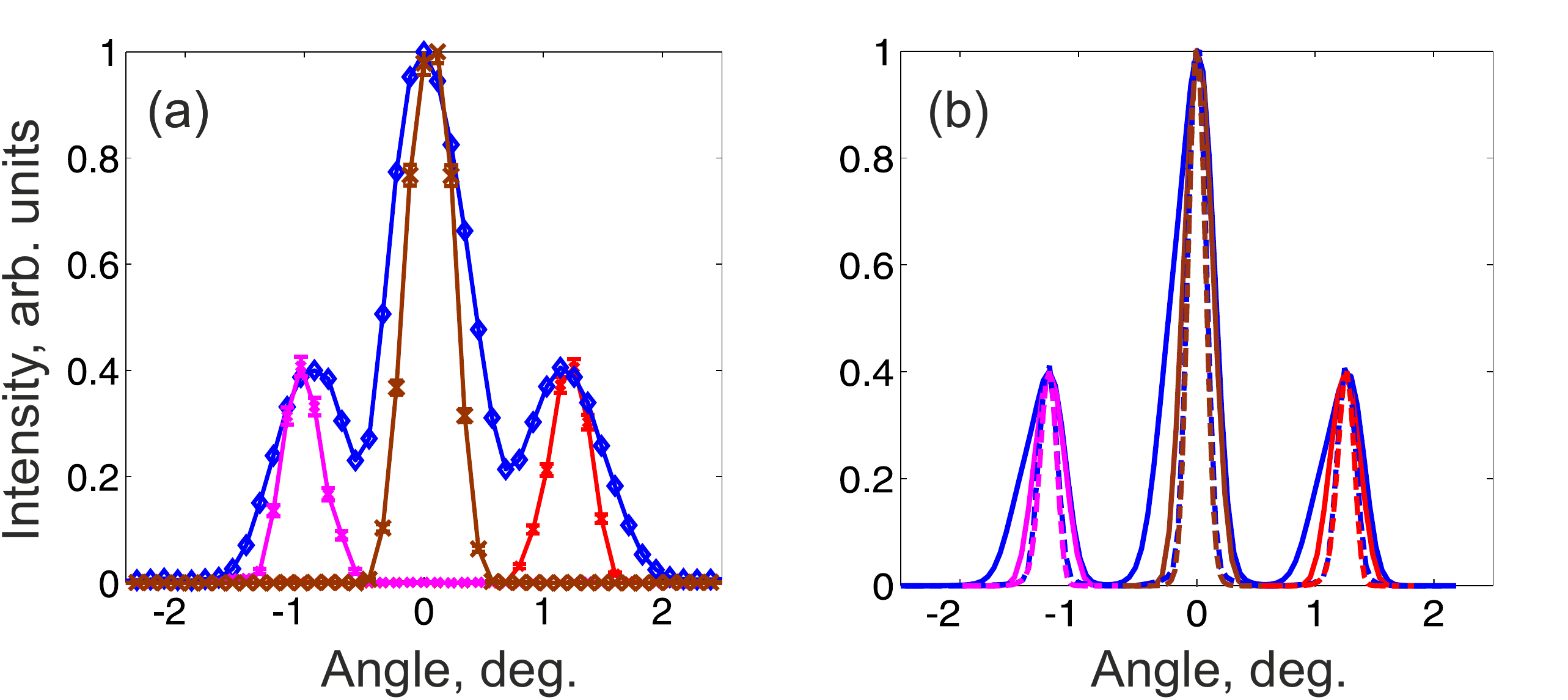}
  \caption{Measured (a) and calculated (b) anglular distributions of SPDC for $M$=3. Blue points and lines are for single count rates and red, brown, and magenta are for two-photon count coincidence rates.
  Solid lines in panel (b) are calculated for parameters of the experiment and dash ones reflect the  case of the weak pump focusing in both $x$ and $y$ directions and narrow slits of sizes $0.2\times0.5$~mm in both directions.}\label{fig:3pumps}
  \end{center}
  \end{figure}

The measured and calculated angular intensity distributions for SPDC
are shown in Fig.~\ref{fig:3pumps}. Blue lines show single count rate spectra
obtained by scanning a slit in the signal beam. Red, brown and magenta are joint two-photon count rate spectra obtained by scanning a
slit in the signal beam after placing another slit in the idler beam at three different positions of maxima in the single count rate
distribution. Solid lines in panel (b) are calculated for the parameters of the experiment and dash ones assume a weak pump focusing in both $x$ and $y$ directions (FWHM of the field envelope $250\,\mu\rm{m}$) and narrow slits of sizes $0.2\times0.5$~mm in both directions.

The noticeable broadening of the experimental distributions compared to the theoretical ones is due to the large height of the slits, which results in integrating of the spectrum intensity in the $x$ direction (compare with the dash lines in Fig.~\ref{fig:3pumps}b corresponding to the theoretical distributions with very narrow slits). Another reason of the broadening was the effect of the pump beam walkoff, which can be overcome by modulating pump beam shape not only in $y$ but both in $x$ and $y$ directions. The usage of frequency filters of smaller width will also result in a narrower SPDC distribution. We explain the difference in conditional and unconditional experimental spectra by misalignment in the slit positions.

\begin{figure}[]
  \begin{center}
  \includegraphics[scale=0.5]{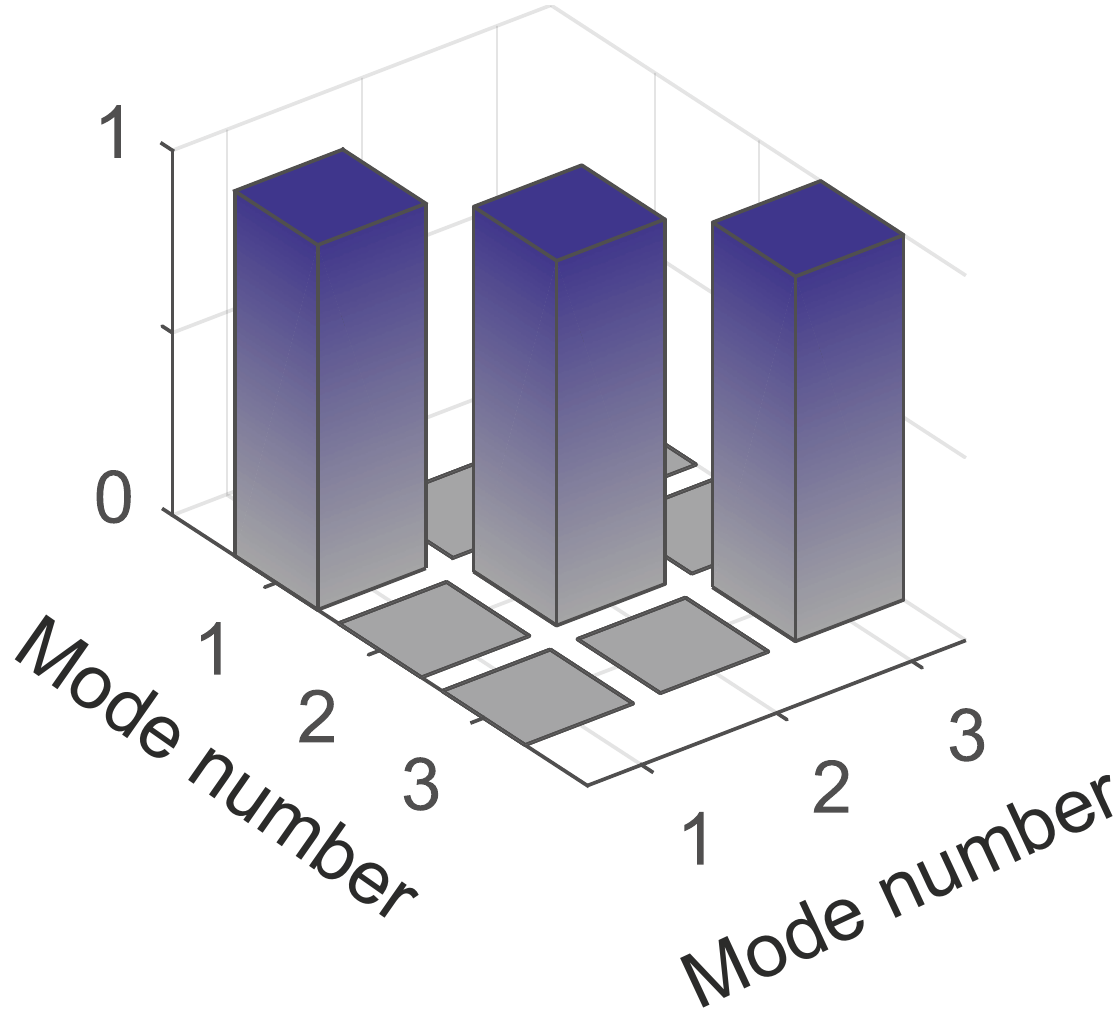}
  \caption{Intensity correlations calculated for each pair of modes for the parameters of the experiment.}\label{fig:hist}
  \end{center}
  \end{figure}

In our approach the purity of the SPDC distributions produced by each of three interfering Gaussian pump beams can theoretically reach almost $1$ as it equals to the inverse of the Schmidt number \cite{Zhan2015}, with cross-correlations less then $10^{-41}$ (see Fig.~\ref{fig:hist}).
This value was calculated as a normalized square of an overlapping area for theoretical angular distributions of each pair of modes.

\section{Conclusion}
Here we have demonstrated a simple method
of achieving separated Schmidt modes in the angular spectrum of SPDC. The method is based on spatially modulating the pump intensity distribution on the nonlinear crystal.
 Another advantage of our approach is that an arbitrary shape of a SPDC angular distribution can be achieved due to the usage of a spatial light modulator at the stage of pump preparation.
\section{Acknowledgments}
We thank Angela Perez for the help with the SLM.

\section{Funding}
This work was supported by the Russian Foundation of Basic Research  (project 17-02-00790 A), Foundation for the Advancement of Theoretical Physics and Mathematics “BASIS” (Postdoc 17-13-334-1), and the Program no. 0066-2019-0005  of the Ministry of Science and Higher Education of Russia for Valiev Institute of Physics and Technology of RAS.

\section{\refname}
\renewcommand{\refname}{References}
\bibliographystyle{unsrt}

\bibliography{library07052019}

\begin{thebibliography}{10}

\bibitem{stenholm1998}
Stig Stenholm and Patrick~J Bardroff.
\newblock {Teleportation of N -dimensional states}.
\newblock {\em Phys. Letters Ser. A.}, 58(6):4373--4376, 1998.

\bibitem{Wang2005}
Chuan Wang, Fu-Guo Deng, Yan-Song Li, Xiao-Shu Liu, and Gui~Lu Long.
\newblock {Quantum secure direct communication with high-dimension quantum
  superdense coding}.
\newblock {\em Physical Review A}, 71(4):44305, 2005.

\bibitem{Zhao2012a}
Rui-Tong Zhao, Qi~Guo, Li~Chen, Hong-Fu Wang, and Shou Zhang.
\newblock {Quantum superdense coding based on hyperentanglement}.
\newblock {\em Chinese Physics B}, 21(8):080303, aug 2012.

\bibitem{Bourennane2001}
Mohamed Bourennane, Anders Karlsson, and Gunnar Bj{\"{o}}rk.
\newblock {Quantum key distribution using multilevel encoding}.
\newblock {\em Physical Review A}, 64(1):1--5, 2001.

\bibitem{Bruss2002}
D~Bruss and C~Macchiavello.
\newblock {Optimal Eavesdropping in Cryptography with Three-Dimensional Quantum
  States}.
\newblock {\em Physical Review Letters}, 88(12):1--4, 2002.

\bibitem{Groblacher2006}
Simon Gr{\"{o}}blacher, Thomas Jennewein, Alipasha Vaziri, Gregor Weihs, and
  Anton Zeilinger.
\newblock {Experimental quantum cryptography with qutrits}.
\newblock {\em New Journal of Physics}, 8(5):75--75, may 2006.

\bibitem{Avella2014}
A.~Avella, M.~Gramegna, A.~Shurupov, G.~Brida, M.~Chekhova, and M.~Genovese.
\newblock {Separable Schmidt modes of a nonseparable state}.
\newblock {\em Physical Review A}, 89(2):023808, feb 2014.

\bibitem{Franson1989}
J~D Franson.
\newblock {Bell Inequality for Position and Time}.
\newblock {\em Physical Review Letters}, 62(19), 1989.

\bibitem{Bechmann-Pasquinucci2000a}
H~Bechmann-Pasquinucci and W~Tittel.
\newblock {Quantum cryptography using larger alphabets}.
\newblock {\em Physical Review A}, 61(6):62308, 2000.

\bibitem{th2004}
R~Thew, A.~Ac{\'{i}}n, H~Zbinden, and N~Gisin.
\newblock {Experimental realization of entangled qutrits for quantum
  communication}.
\newblock {\em Quantum Information {\&} Computation}, 4(93):093--101, 2004.

\bibitem{DeRiedmatten2004}
Hugues de~Riedmatten, Ivan Marcikic, Valerio Scarani, Wolfgang Tittel, Hugo
  Zbinden, and Nicolas Gisin.
\newblock {Tailoring photonic entanglement in high-dimensional Hilbert spaces}.
\newblock {\em Physical Review A}, 69(5):050304, may 2004.

\bibitem{Stucki2005}
Damien Stucki, Hugo Zbinden, and Nicolas Gisin.
\newblock {A Fabry-Perot like two-photon interferometer for high-dimensional
  time-bin entanglement}.
\newblock {\em Journal of Modern Optics}, 52(18), feb 2005.

\bibitem{OSullivan-Hale2005}
Malcolm O'Sullivan-Hale, Irfan {Ali Khan}, Robert Boyd, and John Howell.
\newblock {Pixel Entanglement: Experimental Realization of Optically Entangled
  d=3 and d=6 Qudits}.
\newblock {\em Physical Review Letters}, 94(22):220501, jun 2005.

\bibitem{Neves2005}
Leonardo Neves, G.~Lima, J.~{Aguirre G{\'{o}}mez}, C.~Monken, C.~Saavedra, and
  S.~P{\'{a}}dua.
\newblock {Generation of Entangled States of Qudits using Twin Photons}.
\newblock {\em Physical Review Letters}, 94(10):100501, mar 2005.

\bibitem{Bartuskova2006}
Lucie Bartuskova, Anton{\'{i}}n Cernoch, Radim Filip, Jarom{\'{i}}r Fiurasek,
  Jan Soubusta, and Miloslav Dusek.
\newblock {Optical implementation of the encoding of two qubits to a single
  qutrit}.
\newblock {\em Physical Review A}, 74(022325), aug 2006.

\bibitem{Walborn2006}
S.~Walborn, D.~Lemelle, M.~Almeida, and P.~Ribeiro.
\newblock {Quantum Key Distribution with Higher-Order Alphabets Using Spatially
  Encoded Qudits}.
\newblock {\em Physical Review Letters}, 96(9):090501, mar 2006.

\bibitem{Rossi2009}
Alessandro Rossi, Guiseppe Vallone, Andrea Chiuri, Francesco {De Martini}, and
  Paolo Mataloni.
\newblock {Multi-path entanglement of two photons}.
\newblock {\em Physical Review Letters}, 102(153902), 2009.

\bibitem{Vaziri2002}
Alipasha Vaziri, Gregor Weihs, and Anton Zeilinger.
\newblock {Experimental Two-Photon, Three-Dimensional Entanglement for Quantum
  Communication}.
\newblock {\em Physical Review Letters}, 89(24):240401, nov 2002.

\bibitem{Malik2014}
Mehul Malik, Mohammad Mirhosseini, Martin P~J Lavery, Jonathan Leach, Miles~J
  Padgett, and Robert~W Boyd.
\newblock {Direct measurement of a 27-dimensional orbital-angular-momentum
  state vector.}
\newblock {\em Nature communications}, 5(3115), jan 2014.

\bibitem{Gisin2002}
Nicolas Gisin, Gr{\'{e}}goire Ribordy, Wolfgang Tittel, and Hugo Zbinden.
\newblock {Quantum cryptography}.
\newblock {\em Reviews of Modern Physics}, 74(1):145--195, 2002.

\bibitem{Harris2017}
Nicholas~C. Harris, Gregory~R. Steinbrecher, Jacob Mower, Yoav Lahini, Mihika
  Prabhu, Tom Baehr-Jones, Michael Hochberg, Seth Lloyd, and Dirk Englund.
\newblock {Bosonic transport simulations in a large-scale programmable
  nanophotonic processor}.
\newblock {\em Nature Photonics}, 11:447--452, 2017.

\bibitem{Monken1998}
C~H Monken, P~H~S Ribeiro, and S~P{\'{a}}dua.
\newblock {Transfer of angular spectrum and image formation in spontaneous
  parametric down-conversion}.
\newblock {\em Physical Review A}, 57(4):3123--3126, 1998.

\bibitem{Banaszek2003}
A~B U'Ren, K~Banaszek, and I~A Walmsley.
\newblock {Photon engineering for quantum information processing}.
\newblock {\em Quantum Information {\&} Computation}, 3(October):480--502,
  2003.

\bibitem{law2004}
C~Law and J~Eberly.
\newblock {Analysis and Interpretation of High Transverse Entanglement in
  Optical Parametric Down Conversion}.
\newblock {\em Physical Review Letters}, 92(12):1--4, 2004.

\bibitem{Fedorov2018}
M~V Fedorov.
\newblock {High resource of azimuthal entanglement in terms of Cartesian
  variables of noncollinear biphotons}.
\newblock {\em Phys. Rev. A}, 97(012319):1--7, 2018.

\bibitem{Fedorov2015a}
M~V Fedorov.
\newblock {Schmidt decomposition for non-collinear biphoton angular wave
  functions}.
\newblock {\em Physica Scripta}, 90(074048):1--7, 2015.

\bibitem{Fedorov2009}
M~V Fedorov, Yu~M Mikhailova, and P~A Volkov.
\newblock {Gaussian modelling and Schmidt modes of SPDC biphoton states}.
\newblock {\em Journal of Physics B: Atomic, Molecular and Optical Physics},
  42(17):175503, 2009.

\bibitem{Straupe2011}
S~S Straupe, D~P Ivanov, A~A Kalinkin, I~B Bobrov, and S~P Kulik.
\newblock {Angular Schmidt modes in spontaneous parametric down-conversion}.
\newblock {\em Physical Review A}, 83(6):60302, 2011.

\bibitem{Zhou2017}
Yiyu Zhou, Mohammad Mirhosseini, Dongzhi Fu, Jiapeng Zhao, Seyed
  Mohammad~Hashemi Rafsanjani, Alan~E Willner, and Robert~W Boyd.
\newblock {Sorting photons by radial quantum number}.
\newblock {\em Physical Review Letters}, 119(263602), 2017.

\bibitem{Ruffato2018}
Gianluca Ruffato, Marcello Girardi, Michele Massari, Erfan Mafakheri, Bereneice
  Sephton, Pietro Capaldo, Andrew Forbes, and Filippo Romanato.
\newblock {OPEN A compact diffractive sorter for high-resolution demultiplexing
  of orbital angular momentum beams}.
\newblock {\em Scientific reports}, 8(10248):1--12, 2018.

\bibitem{Zhou2019}
Yiyu Zhou, Mohammad Mirhosseini, Stone Oliver, Jiapeng Zhao, Seyed
  Mohammad~Hashemi Rafsanjani, Martin P~J Lavery, Alan Willner, and Robert~W.
  Boyd.
\newblock {Using all transverse degrees of freedom in quantum communications
  based on a generic mode sorter}.
\newblock {\em Optics Express}, 27(7):10383--10394, 2019.

\bibitem{Ghosh2018}
Debadrita Ghosh, Thomas Jennewein, Piotr Kolenderski, and Urbasi Sinha.
\newblock {Spatially correlated photonic qutrit pairs using pump beam
  modulation technique}.
\newblock {\em OSA Continuum}, 1(3):996--1011, 2018.

\bibitem{Bolduc2013}
Eliot Bolduc, Nicolas Bent, Enrico Santamato, Ebrahim Karimi, and Robert~W
  Boyd.
\newblock {Exact solution to simultaneous intensity and phase encryption with a
  single phase-only hologram.}
\newblock {\em Optics letters}, 38(18):3546--9, sep 2013.

\bibitem{Fedorov2004}
M~V Fedorov, M~A Efremov, A~E Kazakov, K~W Chan, C~K Law, and J~H Eberly.
\newblock {Packet narrowing and quantum entanglement in photoionization and
  photodissociation}.
\newblock {\em Physical Review A}, 69(5):52117, 2004.

\bibitem{Zhan2015}
Mengying Zhan, Qichao Sun, Tong Xiang, and Xianfeng Chen.
\newblock {Generation of high spectral purity photon-pairs with MgO-doped
  periodically poled lithium niobate}.
\newblock {\em Laser Physics}, 25(125203), 2015.

\end{thebibliography}

\end{document}